\documentclass[acmsmall,screen,nonacm]{acmart}

\startPage{1}

\setcopyright{rightsretained}
\copyrightyear{2023}           %% If different from \acmYear

\usepackage{booktabs}   %% For formal tables:
\usepackage{subcaption} %% For complex figures with subfigures/subcaptions
\usepackage{hyperref}

\usepackage{graphics}

\usepackage{listings}

\begin{document}

%% Title information
\title[Schema Change Challenge]{Live \& Local Schema Change: Challenge Problems}
%\subtitle{Subtitle}

%% Author information
%% Contents and number of authors suppressed with 'anonymous'.
%% Each author should be introduced by \author, followed by
%% \authornote (optional), \orcid (optional), \affiliation, and
%% \email.
%% An author may have multiple affiliations and/or emails; repeat the
%% appropriate command.
%% Many elements are not rendered, but should be provided for metadata
%% extraction tools.

%% Author with single affiliation.
\author{Jonathan Edwards}
%\authornote{with author1 note}          %% \authornote is optional;
                                        %% can be repeated if necessary
\orcid{0000-0003-1958-7967}             %% \orcid is optional
\affiliation{
%  \position{Position1}
%  \department{Department1}              %% \department is recommended
  \institution{independent}            %% \institution is required
%  \streetaddress{Street1 Address1}
  \city{Boston}
  \state{MA}
%  \postcode{Post-Code1}
  \country{US}                    %% \country is recommended
}
\email{jonathanmedwards@gmail.com}          %% \email is recommended

\author{Tomas Petricek}
\orcid{0000-0002-7242-2208}             %% \orcid is optional
\affiliation{
%  \department{Distributed and Dependable Systems}             %% \department is recommended
  \institution{Charles University}           %% \institution is required
%  \streetaddress{Street2a Address2a}
  \city{Prague}
%  \state{State2a}
%  \postcode{Post-Code2a}
  \country{CZ}                   %% \country is recommended
}
\email{tomas@tomasp.net}         %% \email is recommended

\author{Tijs van der Storm}
\orcid{0000-0001-8853-7934}             %% \orcid is optional
\affiliation{
%  \department{Distributed and Dependable Systems}             %% \department is recommended
  \institution{Centrum Wiskunde \&\ Informatica (CWI)}           %% \institution is required
%  \streetaddress{Street2a Address2a}
  \city{Amsterdam}
%  \state{State2a}
%  \postcode{Post-Code2a}
  \country{NL}                   %% \country is recommended
}
\affiliation{
%  \department{Distributed and Dependable Systems}             %% \department is recommended
  \institution{University of Groningen}           %% \institution is required
%  \streetaddress{Street2a Address2a}
  \city{Groningen}
%  \state{State2a}
%  \postcode{Post-Code2a}
  \country{NL}                   %% \country is recommended
}

\email{storm@cwi.nl}         %% \email is recommended

%% Abstract
%% Note: \begin{abstract}...\end{abstract} environment must come
%% before \maketitle command
\begin{abstract}

Schema change is an unsolved problem in both live programming and local-first software. We include in schema change any change to the expected shape of data, whether that is expressed explicitly in a database schema or type system, or whether those expectations are implicit in the behavior of the code. Schema changes during live programming can create a mismatch between the code and data in the running environment. Similarly, schema changes in local-first programming can create mismatches between data in different replicas, and between data in a replica and the code colocated with it. In all of these situations the problem of schema change is to migrate or translate existing data in coordination with changes to the code.

This paper contributes a set of concrete scenarios involving schema change that are offered as challenge problems to the live programming and local-first communities. We hope that these problems will spur progress by providing concrete objectives and a basis for comparing alternative solutions.

\end{abstract}

%% 2012 ACM Computing Classification System (CSS) concepts
%% Generate at 'http://dl.acm.org/ccs/ccs.cfm'.
%% End of generated code

%% Keywords comma separated list
%% \keywords are mandatory in final camera-ready submission
%\keywords{schema change, live programming, local-first software}

%% \maketitle
%% Note: \maketitle command must come after title commands, author
%% commands, abstract environment, Computing Classification System
%% environment and commands, and keywords command.
\maketitle

\section{Introduction}

Schema change won't go away. Changing requirements and changing code lead to changing the schema of a database. Schema change, also called schema migration, is the problem of migrating existing data from the old schema to the new. This often involves custom migration programs or specialized Domain Specific Languages. The migration must be carefully coordinated with upgrading the application code that assumes the new schema. Schema change on database servers is often delegated to Database Administrators and DevOps. Live programming~\cite{tanimoto90,hancock03} and local-first software~\cite{localfirst} both forgo centralized database administration, but still must deal with schema change, and must do so on the fly. Recently \citet{Cambria} spotlighted these problems and proposed an approach using lenses~\cite{Foster2007}, but otherwise there has been surprisingly little research. To promote further progress we offer a set of challenge problems to the live programming and local-first communities, inviting them to propose and compare solutions.

Live programming seeks to erase the boundary between editing and running programs. In order to do so program data must be kept around while the program is being edited. Classic Lisp and Smalltalk systems integrated code and data into a persistent \textit{image}~\cite{Sandewall78, Goldberg80}. An interactive shell or \textit{Read Eval Print Loop (REPL)}~\cite{Deutsch64} is more transient than an image, but still builds up a context of data over long-lived programming sessions the loss of which disrupts the programmer's workflow. In all of these environments programs eventually get changed to create and expect data in a form incompatible with extant data. This happens whether or not the form of the data is specified explicitly in a type system or database schema, or whether it is left implicit in the behavior of the code. Some languages can tolerate a larger set of such changes, most notably Smalltalk which will automatically insert and delete members in existing instances when a class definition is changed~\cite[pp.252-272]{Goldberg80}.\footnote{Gemstone turns the Smalltalk image into a production-quality database and accordingly provides a sophisticated schema change API~\cite{Gemstone}.} But there is still a large class of changes that create incompatibilities with existing data, whether it is inside an image, programming environment, REPL, or an external database. Some live programming environments generate data with unit tests, but that only shifts the problem of schema change to adapting those tests. One way or another, stopping everything to manually deal with schema change contradicts the goals of live programming. Schema change won't go away.

Local-first software faces related problems. Its goal is to empower users by moving code and data from the cloud to the user's own devices. Distributed programming techniques like Convergent Replicated Data Types (CRDTs)~\cite{Shapiro11} are used to coordinate data changes peer-to-peer. Unfortunately these techniques so far have not addressed schema change nor code deployment. The traditional techniques of schema change used in centrally managed databases are complicated by the distributed and intermittently connected nature of local-first data. Schema change won't go away.

In the following sections we present a series of challenge problems dealing with schema change in the context of live programming and local-first software. These problems are necessarily expressed using established idioms or conventions, but nevertheless we welcome solutions that translate the spirit of the problem into other contexts.

\section{Extract Entity}

\subsection{Context}
In this challenge we consider a typical example of how the design of a database evolves as new needs are discovered. Much has been written on the subject of data modeling methodologies and the theory of relational database normalization~\cite{Molina08}. But that work often focuses on determining the correct data model up front before the system goes live, which would indeed be the optimal solution if only we were omniscient. The Extract Entity challenge confronts the practical reality of evolving a relational data model in flight.

\subsection{Example}
The Acme Corporation needs to record orders from various customers for various products. The simplest implementation is a spreadsheet with a row for each order and columns for information about the customer and product. It might look like Table \ref{table1}.

\begin{table}[h!]
\centering
  \begin{tabular}{ |l|c|c|l|l|}
   \hline
   Item & Quantity & Ship Date & Customer name & Customer address \\
   \hline \hline
   Anvil & 1 & 2/3/23 & Wile E Coyote & 123 Desert Station \\
   \hline
   Dynamite & 2 & & Daffy Duck & White Rock Lake \\
   \hline
   Bird Seed & 1 & & Wile E Coyote & 123 Desert Station \\
   \hline
  \end{tabular}
\caption{Orders}
\label{table1}
\end{table}

The shipping department filters this table on blank ship dates to see what they need to ship. But after a while the orders department realizes they are wasting effort duplicating the address for new order from an old customer. And when the customer's address changes they have to go back and edit all of their orders. What is needed is two tables, one with orders that links to one with customers, like Tables \ref{table2} and \ref{table3}.

\begin{table}[h!]
  \centering
    \begin{tabular}{ |l|l|}
      \hline
      Name & Address \\
      \hline \hline
      Wile E Coyote & 123 Desert Station \\
      \hline
      Daffy Duck & White Rock Lake \\
      \hline
    \end{tabular}
  \caption{Customers}
  \label{table2}
\end{table}

\begin{table}[h!]
  \centering
    \begin{tabular}{ |l|c|c|l|}
     \hline
     Item & Quantity & Ship Date & Customer name \\
     \hline \hline
     Anvil & 1 & 2/3/23 & Wile E Coyote \\
     \hline
     Dynamite & 2 & & Daffy Duck  \\
     \hline
     Bird Seed & 1 & & Wile E Coyote \\
     \hline
    \end{tabular}
  \caption{Orders linking to Customers}
  \label{table3}
\end{table}

We often encounter this sort of design change when we realize that attributes of one type of entity should actually belong to a new type of entity that will be associated with the first one. Often the motivation is to centralize changes to those attributes in one place. We call this operation \textit{Extract Entity}.

In a SQL database, \texttt{Customers.Name} would be a primary key and \texttt{Orders.Customer} would be a foreign key. But before \texttt{Customers.Name} can be made a primary key the two occurrences of \texttt{Wile E Coyote} must be deduplicated one way or another~\cite{dedupe}. Given the dominance of SQL databases it is not surprising that there are many tools for performing schema change on them~\cite{RailsMigrations, liquibase}. We do find it surprising how low-level these tools are. There is support for the basic SQL DDL commands like adding and dropping tables, columns, or constraints, but no higher level operations that would help solve our problem. One is left to write custom SQL, which can become difficult if the migration needs to be performed online without bring down the database. There is a rich theory of relational database \textit{normalization} based on \textit{functional dependencies}~\cite{Molina08} but that theory has not been implemented into schema change operations in databases -- all you get is SQL. \citet{ambler06} define higher-level patterns of schema change like \textit{Split Table} that sketch SQL implementations but don't provide everything required for Extract Entity. These difficulties fairly beg for new linguistic abstractions, though programming patterns and APIs are reasonable pragmatic solutions to our challenge.

The Extract Entity schema change is straightforward in a spreadsheet. The \texttt{Customer Name} and \texttt{Customer Address} columns are copied and pasted into a new Customers table. The \textit{Remove Duplicates} command is used to deduplicate the customers. To preserve the original view for the shipping department the \texttt{Customer Address} column is redefined with a \texttt{VLOOKUP} formula to pull the address out of the customer table based on the customer's name. A spreadsheet won't maintain referential integrity between the two tables, but end-user database tools can do that~\cite{airtable, notion}.

\subsection{Challenge}
The Extract Entity challenge for live programming is to provide interactive operations in the programming environment that will perform the schema change on live code and data. The data can be in any form desired, including relations, objects, or JSON. The code can be a simple CRUD UI that also provides the shipping department its view of unshipped orders. The central problem is to coordinate the schema change with code edits/refactorings to keep the system executing live and correctly. Solutions should try to minimize:

\begin{enumerate}
  \item The number and complexity of commands or UI affordances that must be introduced.
  \item Interruption to live execution of the code.
\end{enumerate}

The Extract Entity challenge for local-first software is similar to the live programming challenge except that the context is an application that offers similar functionality with peer-to-peer collaboration. In this context the schema change does not need to be performed interactively — a developer can take some time to build, test, and package a solution, perhaps using an API or DSL. The hard part comes when the schema change is to be deployed across all the replicas. This deployment must synchronize code and data upgrades (or decouple them as in Cambria~\cite{Cambria}). The deployment must also deal with migrating ``in flight'' operations so that all replicas converge on the same state without data loss, and ideally without centralized coordination. A radical approach could try to incorporate schema change operations into the underlying datastore/CRDT itself and integrate code as well, but layered solutions are welcome too.

\subsection{Goals}
Solutions should try to minimize:

\begin{enumerate}
  \item The need for users to manually intervene.
  \item Possibilities that developer-written code is subtly incorrect in edge cases, preferably by reducing the need for developer-written code.
  \item The need for a centralized coordinator.
  \item How long replicas might be partitioned until resynchronization is acquired.
\end{enumerate}

\section{Divergence Control}

\subsection{Context}
Schema change is usually concerned with migrating old data into a new schema. But sometimes we also want to migrate in the reverse direction: new data into an old schema. \citet{Cambria} discuss various ways that code changes can require bidirectional migration. Instead in this problem we focus on changes made by end-users. Users frequently make copies of structured documents like spreadsheets and then diverge them by altering both data content and schema. In the case of spreadsheets schema changes include rearrangements to the structure of rows and columns as well as changes to formulas. The problem is that the users want to transfer such data and schema changes between these divergent copies, and they want to pick and choose which of these changes to transfer. In practice such transfer is done manually through copy \& paste. \citet{Basman19} has documented an ecology of emailed spreadsheets. \citet{Burnett14} distilled field observations of these practices in the story of Frieda, which we quote here:

\begin{quotation}

For example, consider ``Frieda'', an office manager in charge of her department's budget tracking. (Frieda was a participant in a set of interviews with spreadsheet users that the first author conducted. Frieda is not her real name.) Every year, the company she works for produces an updated budget tracking spreadsheet with the newest reporting requirements embedded in its structure and formulas. But this spreadsheet is not a perfect fit to the kinds of projects and sub-budgets she manages, so every year Frieda needs to change it. She does this by working with four variants of the spreadsheet at once: the one the company sent out last year (we will call it Official-lastYear), the one she derived from that one to fit her department's needs (Dept-lastYear), the one the company sent out this year (Official-thisYear), and the one she is trying to put together for this year (Dept-thisYear).

Using these four variants, Frieda exploratively mixes reverse engineering, reuse, programming, testing, and debugging, mostly by trial-and-error. She begins this process by reminding herself of ways she changed last year's by reverse engineering a few of the differences between Official-lastYear and Dept-lastYear. She then looks at the same portions of Official-thisYear to see if those same changes can easily be made, given her department's current needs.

She can reuse some of these same changes this year, but copying them into Dept-thisYear is troublesome, with some of the formulas automatically adjusting themselves to refer to Dept-lastYear. She patches these up (if she notices them), then tries out some new columns or sections of Dept-thisYear to reflect her new projects. She mixes in ``testing'' along the way by entering some of the budget values for this year and eyeballing the values that come out, then debugs if she notices something amiss. At some point, she moves on to another set of related columns, repeating the cycle for these. Frieda has learned over the years to save some of her spreadsheet variants along the way (using a different filename for each), because she might decide that the way she did some of her changes was a bad idea, and she wants to revert to try a different way she had started before.
\end{quotation}

\subsection{Example}
The Divergence Control challenge instantiates the problem of divergent state with the example from the Extract Entity challenge. Every month the orders department sends its spreadsheet to the accounting department, which adds the data to a version of the spreadsheet that it has customized for financial tracking. But when the orders department migrates its spreadsheet to extract out customers the accounting department does not want to conform. They could manually convert incoming data in the new schema into their variant of the old schema. But they shouldn't have to. It should be possible to run new data through the schema change in reverse, converting it back into the format the accounting department is used to ingesting. Note that this is not a matter of synchronizing the divergent spreadsheets -- they maintain differently evolved schema.

\subsection{Challenge}
What is needed is a user-friendly mechanism for transferring data changes from one schema to another bidirectionally. In one direction we want to transfer changes made in the new schema into a variant of the old schema. The opposite direction might be needed, for example, if the accounting department makes corrections to customer addresses, which ought to be pushed through into the orders department's new schema.

Bidirectional transfer of changes between divergent copies is similar in some ways to source code version control as in Git~\cite{ProGit}. There is \textit{forking} of long-lived divergent copies. We want to \textit{diff} these forks to see exactly how they have diverged. We want to partially \textit{merge} them by \textit{cherry picking} certain differences. Yet there are also many dissimilarities with Git: our data is more richly structured than lines of text; our schema changes are higher-level transformations on these structures than inserting and deleting characters; and we expect that end-users be able to understand it~\cite{gitless}. The Divergence Control challenge is in a sense to provide ``version control for schema change'' meeting these criteria, with the key technical challenge being the ability to transfer selected differences bidirectionally through schema changes as independently as possible.

\subsection{Goals}
This challenge invites a change of perspective in both live programming and local-first software. Live programming must move from being a solitary activity to a collaborative workflow, and one where the collaboration is not just on editing source code but on live integrated code and data, as in the Smalltalk/Lisp images of old.\footnote{If only we had collaborative and deployable images back in the 90's when Smalltalk had a shot at the mainstream!} For its part, local-first software must move from automatically converging data replicas to also interactively managing long-term divergence. That implies either application state is being directly exposed to the user, or the application code is using an API to present version control affordances. We encourage both communities to think outside the box of Git, which has proven to baffle not only end-users but also a substantial fraction of developers.

\section{Conference Organizer: Merging structural document edits}

\subsection{Context}
In this challenge, we consider a document-based format akin to that used, for example, by Notion\cite{notion}. We focus on schema change applied to data, but the document-based format can also be extended to include programming as done, for example in Boxer\cite{Boxer}. We assume that the document can be concurrently edited by multiple users in the same fashion as in local-first software, i.e., users may edit a local copy that then needs to be synchronized with changes made by other users. The edits to the document can be done in a rich-text editor, or by using a more structure-aware structure editor that can, for example, record changes as a sequence of high-level edit operations.

\subsection{Example}
In this challenge, we consider the problem of collaborative conference planning. A number of co-organizers are planning a conference that will feature talks by several invited speakers. The organizers need to agree on a list of speakers to invite, coordinate who contacts whom and track the acceptance of invitations. They start with a document shown in Figure~\ref{fig:conf-orig}.

\begin{figure}[h!]
\fbox{\parbox{24em}{
\textbf{PROGRAMMING CONFERENCE 2023}\\
\textbf{Invited speakers}
\begin{itemize}
 \item Adele Goldberg, adele@xerox.com
 \item Margaret Hamilton, hamilton@mit.com
 \item Betty Jean Jennings, betty@rand.com
\end{itemize}
}}
\caption{Conference organization. Initial state of the document.}
\label{fig:conf-orig}
\end{figure}

\begin{figure}
\fbox{\parbox{24em}{
\textbf{PROGRAMMING CONFERENCE 2023}\\
\textbf{Invited speakers}
\begin{itemize}
 \item Ada Lovelace, lovelace@rsoc.ac.uk
 \item Adele Goldberg, adele@xerox.com
 \item Betty Jean Jennings, betty@rand.com
 \item Margaret Hamilton, hamilton@mit.com
\end{itemize}
}}
\caption{Conference organization. Added a speaker and sorted list.}
\label{fig:conf-add}
\end{figure}

\begin{figure}
\fbox{\parbox{24em}{
\textbf{PROGRAMMING CONFERENCE 2023}\\
\textbf{Invited speakers}

\begin{tabular}{| l | l | l |}
  \hline
  \textbf{Name} & \textbf{Email} & \textbf{Organizer} \\
  \hline \hline
  Adele Goldberg & adele@xerox.com & TP\\
  \hline
  Margaret Hamilton & hamilton@mit.com & JE\\
  \hline
  Betty Jean Jennings & betty@rand.com & JE\\
  \hline
\end{tabular}
}}\caption{Conference organization. Refactored list into a table.}
\label{fig:conf-tab}
\end{figure}

\begin{figure}
\fbox{\parbox{24em}{
\textbf{PROGRAMMING CONFERENCE 2023}\\
\textbf{Invited speakers}

\begin{tabular}{| l | l | l |}
  \hline
  \textbf{Name} & \textbf{Email} & \textbf{Organizer} \\
  \hline \hline
  Ada Lovelace & lovelace@rsoc.ac.uk & \\
  \hline
  Adele Goldberg & adele@xerox.com & TP\\
  \hline
  Betty Jean Jennings & betty@rand.com & JE\\
  \hline
  Margaret Hamilton & hamilton@mit.com & JE\\
  \hline
\end{tabular}
}}\caption{Conference organization. Resulting document after change merging.}
\label{fig:conf-fin}
\end{figure}

\subsection{Challenge}
The challenge is to merge document edits that are done locally and independently by the co-organizers of the conference. Specifically, consider the following two edits:

\begin{enumerate}
\item The first organizer adds an additional speaker to the list and sorts the list of speakers alphabetically by their first name. (Figure~\ref{fig:conf-add})
\item The second organizer refactors the list into a table. They split the single textual value into a name and an email (using a comma as the separator) and add an additional column for tracking which of the organizers should contact the speaker.
\end{enumerate}

The system should be able to merge the changes and produce a final table shown in Figure~\ref{fig:conf-fin}. The resulting table needs to include the additional speaker created by the first organizer, use the order specified by the first organizer and use the format defined by the second organizer. The newly added speaker should be reformatted into the new format. For the newly added ``Organizer'' column, the second user may specify default value or the newly added row may use an empty value.

\subsection{Goals}
A solution to the challenge should strive to satisfy the following goals:

\begin{itemize}
\item \emph{Completeness.} It should be possible to semi-automatically merge any two sequences of edits. In other words, the merging should always be defined, regardless of what edits the users perform.
\item \emph{Commutativity.} The order of edits should not matter. If users $A$ and $B$ perform edits independently starting from the same initial document, the resulting document should be the same if the system treats edits from $A$ as occurring before the edits of $B$ and vice versa.
\item \emph{Conflict resolution.} It may not always be possible to merge conflicting edits. In this case, the system should interactively ask the user for guidance or, possibly, use default behavior specified by the user.
\item \emph{Convergence.} The system does not need to consider the case where one users does not want to adopt changes created by another user. We assume that all users want to eventually see the same final document produced by applying all edits.
\item \emph{Structure editing.} The system may use a structure editor that provides high-level commands for operations such as reordering of elements or refactoring of a list into a table. In other words, we assume that the challenge may not be solvable in a system based on plain text editing of document source code.
\end{itemize}

\subsection{Remarks}
It is worth noting that this challenge does not discuss explicit distinction between the schema (or ``type'') of the document and data (or ``values''). The implementing system may or may not maintain such distinction. In our example, adding a new speaker and sorting the list is arguably a mere change of the data, but the refactoring of a list into a table is a schema change. The system may leverage knowledge about such distinction and, for example, handle the merging of schema-related and data-related changes differently.

\section{Conference Organizer: Merging edits in the presence of code}

\subsection{Context}
Consider the document-based format as in the previous challenge, but extended with the support for computed values. A computed value is a node in the document whose value is specified by a formula akin to those used in spreadsheets. The equations can refer to other nodes in the document, possibly using either an absolute or relative path and may involve a range of built-in functions, for example to count the number of elements or to sum a sequence of numerical values. We now also assume that elements in the document can have IDs that are not displayed in the default view, but may be used to refer to them in equations.

\begin{figure}
\fbox{\parbox{28em}{
\textbf{PROGRAMMING CONFERENCE 2023}\\
\textbf{Invited speakers}
\begin{itemize}
 \item Adele Goldberg, adele@xerox.com
 \item Margaret Hamilton, hamilton@mit.com
 \item Betty Jean Jennings, betty@rand.com
\end{itemize}
\vspace{0.5em}
\textbf{Conference budget}\\
Travel cost per speaker:\\
\hspace*{2em} \$1200 \\
Number of speakers: \\
\hspace*{2em} \texttt{=COUNT(/ul[id='speakers']/li)} \\
Travel expenses:\\
\hspace*{2em} \texttt{=/dl/dd[0] * /dl/dd[1]}
}}
\caption{Conference organization. Initial state of the document.}
\label{fig:conf-budget}
\end{figure}

\begin{figure}
\fbox{\parbox{28em}{
\textbf{PROGRAMMING CONFERENCE 2023}\\
\textbf{Invited speakers}

\begin{tabular}{| l | l | l |}
  \hline
  \textbf{Name} & \textbf{Email} & \textbf{Organizer} \\
  \hline \hline
  Ada Lovelace & lovelace@rsoc.ac.uk & \\
  \hline
  Adele Goldberg & adele@xerox.com & TP\\
  \hline
  Betty Jean Jennings & betty@rand.com & JE\\
  \hline
  Margaret Hamilton & hamilton@mit.com & JE\\
  \hline
\end{tabular}

\vspace{0.5em}
\textbf{Conference budget}\\
Travel cost per speaker:\\
\hspace*{2em} \$1200 \\
Number of speakers: \\
\hspace*{2em} \texttt{=COUNT(/table[id='speakers']/tbody/tr)} \\
Travel expenses:\\
\hspace*{2em} \texttt{=/dl/dd[0] * /dl/dd[1]}
}}\caption{Conference organization. Resulting document after change merging.}
\label{fig:conf-finfin}
\end{figure}

\subsection{Example}
This challenge extends the previous problem. The conference organizers also want to use the document to manage the conference budget. In a very simplified form, one aspect of this is calculating the estimated travel expenses for all the speakers. For this, they will want to add a computed value to the document that counts the number of speakers and multiplies the result by a fixed cost per speaker elsewhere in the document.

The example is shown in Figure~\ref{fig:conf-budget}. The additional section of the document contains one fixed value (travel cost per speaker) and two computed values. The first selects all \texttt{li} elements of a \texttt{ul} element with ID \texttt{speakers} and counts their number. The second computed value then multiplies the constant travel cost per speaker by the number of speakers. Here, we assume the new section uses the HTML definition list \texttt{dl} and the equation selects its first and second \texttt{dd} item, respectively.

\subsection{Challenge}
The challenge is, again, to merge the edits done independently by the co-organizers. Assume two co-organizers start with the same original document shown in Figure~\ref{fig:conf-orig}.
\begin{enumerate}
    \item The first co-organizer performs the edits outlined previously that result in the document shown in Figure~\ref{fig:conf-fin}, i.e., edits the speakers and refactors the structure to use a table.
    \item The second co-organizer adds the budget calculation as shown in Figure~\ref{fig:conf-budget}.
\end{enumerate}
As before, the system should be able to merge the edits made to the document. Crucially, this requires that the refactoring of the document structure done in (1), is also reflected in the code of the equations added in (2). As shown in Figure~\ref{fig:conf-finfin}, in this specific case, the system changes the equation \texttt{COUNT(/ul[id='speakers']/li)} to \texttt{COUNT(/table[id='speakers']/tbody/tr)}. This is the case because the refactoring consists of four steps that each transform the document and also need to update the code of the equation accordingly:
\begin{itemize}
    \item[--] Change the type of the \texttt{ul} element to \texttt{table}
    \item[--] Wrap the body of the \texttt{table} element with \texttt{tbody}
    \item[--] Change the type of all children of \texttt{tbody} from \texttt{li} to \texttt{tr}
    \item[--] Further split the body into two \texttt{td} elements
\end{itemize}

\subsection{Goals}
A solution to the challenge should meet the goals of the preceding challenge, i.e., completeness (merging should be always defined), commutativity (order does not matter), conflict resolution (ask for guidance if needed), convergence (everyone eventually wants the same document) and structure editing (the system may capture high-level edits).

\subsection{Extensions}
The basic challenge outlined above focuses primarily on the local-first software scenario, but it can be extended to also apply to the live programming scenario. In particular, when the document contains computed values, those may be evaluated either explicitly by the user or automatically on-the-fly. An edit can then invalidate some of the values (either by changing the data that a computed value depends on, or by changing the equation used to compute the value). A live programming system should be able to detect which computed values are affected by an edit and, either invalidate those (requiring an explicit re-evaluation by the user) or automatically re-evaluate them (and possibly highlight the affected values to inform the user).

\section{Live Modeling Languages Require Run-Time State Migration}
%Related work
%``Slogan'': editing a program is diverging from run-time schema.

\subsection{Context}
In stateful live programming, a change in the program can make the run-time state of the execution out of date, hence we want to migrate the run-time state to keep on running.

One can see a program itself as an instance of a static schema (its AST type), that in turn determines (defines/implies/induces) a run-time schema: the run-time structures of code (e.g., classes, inheritance links, declarations, method definitions etc.), as well as the structure of the run-time state. Whereas the run-time \textit{code} structures do not typically change while using an application, the run-time \textit{state} constantly changes.
Live programming, however, requires reconciling changes to the program with the run-time structures of both the code and the state. This typically means that at a certain point (a quiescent point) during execution, the code structures need to be updated (hot swapped), and the state needs to be \textit{migrated}.

\subsection{Example}
Consider the example of a simple state machine language with on-entry actions, and typed global variables (loosely inspired by the SML language of \citet{vanRozen19}). An example state machine and an excerpt of its abstract syntax schema is shown in Figure~\ref{LST:statemachines}. Depicted on the left, an actual statemachine modeling the opening and closing of a door, with one global boolean variable, \lstinline{isClosed}, which is flipped according to transitioning between the \lstinline{closed} and \lstinline{opened} states. For the sake of brevity, the abstract syntax schema (similar, to e.g., and Ecore metamodel~\cite{EMF}) shown on the right of the figure omits classes for the on-entry actions (\lstinline{Stmt}) and variable types (\lstinline{Type}).

\begin{figure}[t]
\centering
\begin{minipage}[t]{0.4\textwidth}
\begin{lstlisting}[language=java,morekeywords={machine,on,state,var}]
machine Doors

var isClosed: bool = true

state closed
  isClosed := true
  on open => opened

state opened
  isClosed := false
  on close => closed
\end{lstlisting}
\end{minipage}
\hspace*{2pt}\vline
\begin{minipage}[t]{0.4\textwidth}
\begin{lstlisting}[language=java,morekeywords={on}]
  class Machine
    name: string
    states: State*
    vars: Decl*

  class State
    name: string
    onEntry: Stmt*
    transitions: Trans*

  class Trans
    event: string
    target: State

  class Decl
    name: string
    type: Type
  \end{lstlisting}
\end{minipage}
\caption{A statemachine and its abstract syntax schema (omitting \lstinline{Stmt} and \lstinline{Type})}
\label{LST:statemachines}
\end{figure}

In the following we assume that the state machine is executed by an interpreter that process an incoming stream of events, and traversing run-time pointers representing transitions accordingly, executing on-entry actions, where the notion of the current state and the values of the global variables are the collective run-time state of the program. The structure that the interpreter operates on can be described as derived  schema from the abstract syntax scheme shown on the right of Figure~\ref{LST:statemachines}, but specialized for a particular machine, e.g., \lstinline{Doors}.
This schema will have additional fields and associations modeling the run-time data required for execution. For the \lstinline{Doors} statemachine, this means adding the current state to the \lstinline{Machine} type (required for any machine) and the machine specific data fields, namely \lstinline{isClosed}. The run-time schema for \lstinline{Doors} is then the following\footnote{It is possible to represent the global variables at run-time using an untyped table or hashmap, but for the sake of schema migration complexity, the encoding as fields in classes is more instructive.}:
\begin{quote}
\begin{lstlisting}[language=java,morekeywords={on},mathescape=true]
class Machine$^{++}$
  states: State*
  current: State
  isClosed: bool
\end{lstlisting}
\end{quote}
This \textit{extended} class \lstinline{Machine}$^{++}$ also represents state machines, but this time at run time.

\begin{figure}[t]
  \centering
\includegraphics[width=0.5\textwidth]{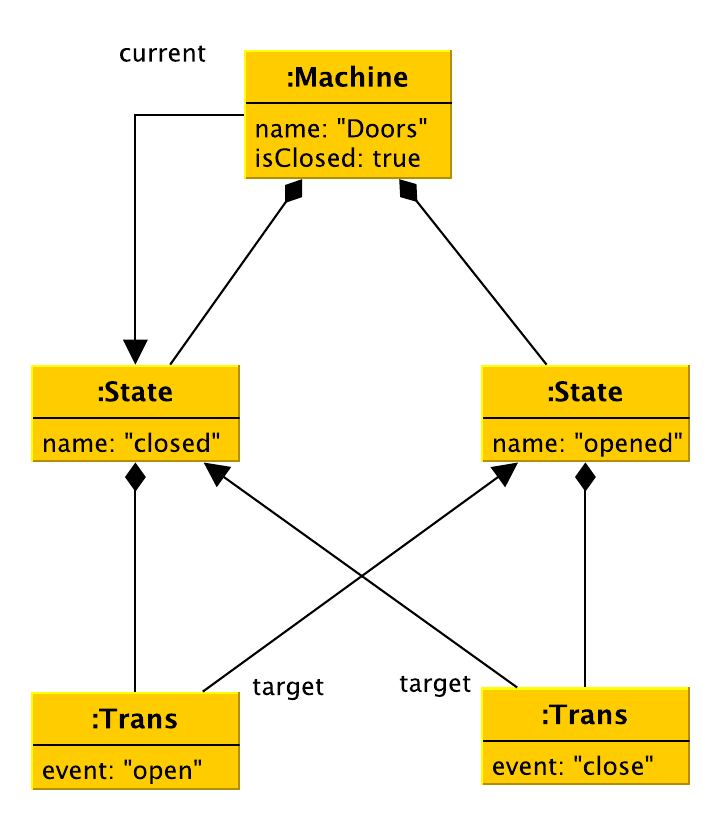}
\caption{UML object diagram of the run-time structures of the Doors statemachine (excluding code objects)}
\label{FIG:doorsRuntime}
\end{figure}

An instance of a running state machine consists of an object graph conforming to the run-time schema of state machines, updating the \lstinline{current} state pointer in response to events. This object graph is graphically depicted in Figure~\ref{FIG:doorsRuntime} as a UML object diagram, excluding the statement objects representing on-entry actions. As the diagram shows, the running machine has a current state (\lstinline{closed}) and the \lstinline{isClosed} field has an actual value (\lstinline{true}). It is instructive to note that the diagram encode both static and dynamic aspects of the state machine language. Live editing the state machine then requires to \textit{patch}~\cite{SemanticDeltas} this run-time structure without shutting it down, possibly requiring migration of run-time state.

\subsection{Challenge}

\begin{table}[t]
  \centering
\begin{tabular}{lll}\toprule
\textsc{Change} & \textsc{What} & \textsc{How}\\\midrule
add & state & simple  \\
remove & state & structurally simple, but heuristic if state is current \\
rename & state & simple\\
add & variable & migrate class and initialize field \\
remove & variable & migrate class (NB: assumes variable is unused)\\
rename & variable & rename in class, preserving value\\
type change & variable & migrate class and convert or reinitialize value \\
add & transition & simple \\
remove & transition & structurally simple, but heuristic for pending events\\
event change & transition & structurally simple, but heuristic for pending events\\
add/remove & statement & hotswap code at quiescent point of interpreter\\
\bottomrule
\end{tabular}
\caption{Possible program changes and how to deal with them at run time}
\label{TBL:changes}
\end{table}

The space of possible (well-formed\footnote{The abstract syntax schema is not expressive enough to define all static invariants of the language, such as: states must by unique by name, references variables in actions must be declared, etc.}) changes to a state machine are summarized in Table~\ref{TBL:changes}.
The first column indicates the change category, the second the affected kind of object, and third a short description of how to deal with the respective change category. Some notes about this table:
\begin{itemize}
\item all rows with ``simple'' in the third column only require a quiescent~\cite{Tranquility} point in the interpreter loop to update the run-time structure (e.g., as shown in Figure~\ref{FIG:doorsRuntime}).
\item removing a state, however, is only \textit{structurally} simple, since removal is easy, but special care needs is required if the subject of removal is the current state. In this case, some heuristic is needed, such as: reject the edit, point the current state to the initial state, or some other strategy (e.g., the nearest state, previous state etc.)
\item everywhere a row mentions ``migrate class'' in the third column, data migration is needed: the run-time objects conforming to the old class must be transformed to instances of the updated class, similar to how Smalltalk migrates objects using \lstinline{become:}.
\item removing a transition or changing the event of a transition potentially has to deal with pending events (e.g., in an event queue) expecting such transitions, since such events are now potentially stale. Strategies to deal with this situation include: simply dropping the events, or requiring that such edits cannot be patched at run time when there are pending events.
\item the type change edit of variable requires a strategy for the current value: either discard and reinitialize, or perform value conversion. For instance, if the type change is from boolean to integer, then true could become 1, and false could become 0.
\item note finally that rename variable is mentioned explicitly as a change: this makes it possible to preserve the run-time value of the variable.
\end{itemize}

\subsection{Goals}

The goals for this challenge are twofold: from the end-user perspective and from the language engineering perspective. A particular challenge from the end-user perspective is to ``do minimal harm'': it is essential for fluid programming experience that the automatically triggered migrations are in a sense as near as possible to the previous application state, to not surprise or confuse the user/developer.
One approach  considered in earlier work~\cite{RuntimeConstraint} employs the constraint solver Z3 to find a ``nearest'' run-time instance compatible with a source change. Nevertheless, the patching of the run-time state should be quick enough so as to not disrupt the programmer experience.

From the language engineering perspective the goal is to employ techniques, formalisms, and tools, to make the construction of such languages easier. The above example state machine DSL is derived from earlier work~\cite{vanRozen19}, where the authors manually implemented the run-time patch operation and concluded that even for such a simple language (simpler even than the example above) it is a complex and error-pone endeavor. Furthermore, the field of software language engineering studies and develops generic and reusable techniques to improve the development of DSLs and programming languages, for instance, in the context of language workbenches~\cite{ERDWEG201524}. The development of live programming languages, however, is currently still out of reach for all existing language workbenches. The aforementioned approach using a constraint solver is an example of such a \textit{language parametric} technique, in that it operates on the (extended) abstract syntax metamodel of a language, and does not assume anything further about the language itself. Another approach is the \textsc{Cascade} metamodeling formalism which has builtin support for run-time patching~\cite{Cascade}. However, further research is needed to design better principled language engineering approaches that solve the problem in a way that is both declarative and fast.

\subsection{Extensions}

While the above example is arguably simple, the problem becomes much more challenging when the programs themselves define data types, classes, records, structures etc. Since possibly many instances (values, objects) of such data types may exist at run-time, these all have to be migrated in such a way that programmer experience is minimally disrupted, and that the invariants of said data types is maintained.
Another extension, tying in with the data-oriented examples above, involves refactorings of data types in a program. Typically a refactoring should be behavior preserving, but can it also preserve run-time data? The minimal example is the consistent variable rename rename in Table~\ref{TBL:changes}, which should not have any effect on run-time state.  A more complex refactoring is described in Section~\ref{SECT:elm}.

\section{Live programming in the context of Elm architecture}
\label{SECT:elm}
\subsection{Context}
Another challenge involving stateful live programming can be drawn from the programming model based on the Elm architecture. In this model, a reactive (web) application is structured in terms of current state and events that affect the state. The implementation then consists of two functions that we may call \texttt{update} and \texttt{render}:

\begin{quote}
\begin{lstlisting}[language=ml,morekeywords={on}]
type State = { .. }
type Event = .. | ..

val update : State -> Event -> State
val render : State -> Html
\end{lstlisting}
\end{quote}

The programming model works as follows:
\begin{itemize}
    \item \texttt{State} represents the entire application state, i.e., everything that the user can work with.
    \item \texttt{Event} represents all events that the user can trigger by interacting with the application.
    \item \texttt{update} is called whenever an event happens. It takes the current state, the event and computes a new state.
    \item \texttt{render} takes the current state and produces a representation of what should be displayed on the screen (e.g., an HTML tree).
\end{itemize}

Traditionally, when Elm applications are developed, they are restarted each time the code is modified and any previous state is discarded. However, a more effective programming model based on live programming would allow live updates to both the code of the two functions and the structure of the two types.

\subsection{Example}

As an example, consider the uninspiring, but well-known, TODO list application. The types that capture the state and events in the application may look as follows:

\begin{quote}
\begin{lstlisting}[language=ml,morekeywords={on}]
type Item = { id : id; title : string; completed : bool }
type State = { items : Item list }
type Event =
  | SetCompleted of id * bool
  | SetTitle of id * string
  | Remove of id
  | Add of string
\end{lstlisting}
\end{quote}

The application state consists of a list of items. Each item has a unique ID alongside with a title and a flag indicating whether it is completed. The events represent edits to the items, deletion and addition. The implementation of the \texttt{update} and \texttt{render} functions is simple and not important for the challenge.

\subsection{Challenge}
Now, imagine that we have a running TODO list application with the above state and events. To test the application, the programmer has already created a number of items and so there is a single value of the \texttt{State} type that represents a current state of the application such as:

\begin{quote}
\begin{lstlisting}[language=ml,morekeywords={on}]
{ items : [
   [ { id = 1; title = "check twitter"; completed = true  }
     { id = 1; title = "Write paper"; completed = false } ] }
\end{lstlisting}
\end{quote}

As above, there are a number of edits to the code and types that the programmer may want to do without restarting the application. Those are summarized in Table~\ref{TBL:elmchanges}. Modifying the code is simple and only requires waiting until the current execution completes. Modifying the \texttt{Event} type is also simple, but it may lead to unused code or missing case in \texttt{update} that needs to be addressed. Finally, modifying \texttt{State} ranges from relatively simple problems (adding a new field) to challenging case when the structure of \texttt{State} is changed.

For example, imagine that the programmer would want to edit the structure of the state and migrate the previous definition of \texttt{State} to the following new definition where individual fields are stored in separate lists:

\begin{quote}
\begin{lstlisting}[language=ml,morekeywords={on}]
type State =
  { ids : id list
    titles string list
    completes : bool list }
\end{lstlisting}
\end{quote}

This representation is semantically equivalent (assuming the lists have the same length) to the original one. It should thus, in principle, be possible to migrate the original state value to a value using the new structure. Moreover, it should, in principle, be also possible to automatically transform the implementations of \texttt{update} and \texttt{render} to work as before, but using the new state structure.

\begin{table}[t]
  \centering
\begin{tabular}{lll}\toprule
\textsc{Change} & \textsc{What} & \textsc{How}\\\midrule
modify & \texttt{render} & simple, but wait until current execution finishes  \\
modify & \texttt{update} & simple, but wait until current execution finishes  \\
add & case to \texttt{Event} & requires adding corresponding case to \texttt{update} \\
remove & case from \texttt{Event} & remove unused code from \texttt{update} \\
add & field to \texttt{State} & migrate state value and initialize field \\
remove & field from \texttt{State} & migrate state (assuming field unused) \\
modify & structure of \texttt{State} & migrate state value and edit code accordingly \\
\bottomrule
\end{tabular}
\caption{Possible program changes and how to deal with them at run time}
\label{TBL:elmchanges}
\end{table}

\subsection{Goals}
As above, the key requirement from the live programming perspective is to “do minimal harm”. In this case, this results in the following goals:

\begin{itemize}
\item When migrating the application state, this needs to be done automatically and the system should strive to produce new state that is as near as possible to the previous state.
\item The solution may rely on structure editing so that the system has access to a high-level logical description of the edits performed by the user.
\item The system needs to handle the case when the type structure diverges from the code structure. This can be addressed in various ways (transform code, add error handlers, etc.) but it is desirable to avoid "breaking" the \texttt{render} function as this would make the new application state impossible to see.
\end{itemize}

\subsection{Remarks}
It is worth noting that the particularly difficult aspect of this challenge, i.e., the case where the structure of state is refactored to an equivalent one, is related to the Extract Entity challenge.

\bibliography{splash23}

%% Appendix
%\appendix
%\section{Appendix}
%Text of appendix \ldots

\end{document}